# A Silicon-on-Insulator Slab for Topological Valley Transport

Xin-Tao He[†], En-Tao Liang[†], Jia-Jun Yuan[†], Hao-Yang Qiu, Xiao-Dong Chen,

Fu-Li Zhao, and Jian-Wen Dong[*]

State Key Laboratory of Optoelectronic Materials and Technologies & School of
Physics, Sun Yat-sen University, Guangzhou 510275, China
[†] These authors contributed equally to this work.
[*] Corresponding author. Email: dongjwen@mail.sysu.edu.cn

**Silicon-on-insulator (SOI) enables for capability improvement of modern information processing systems by replacing some of their electrical counterparts. With the miniaturization of SOI platform, backscattering suppression is one of the central issues to avoid energy loss and signal distortion in telecommunications. Valley, a new degree of freedom, provides an intriguing way for topologically robust information transfer and unidirectional flow of light, in particular for subwavelength strategy that still remains challenge in topological nanophotonics. Here, we realize topological transport in a SOI valley photonic crystal (VPC) slab. In such inversion asymmetry slab, singular Berry curvature near Brillouin zone corners guarantees valley-dependent topological edge states below light cone, maintaining a balance between in-plane robustness and out-of-plane radiation. Topologically robust transport at telecommunication wavelength is observed along two sharp-bend VPC interfaces with a compact size (< 10 μm), showing flat-top high transmission of around 10% bandwidth. Furthermore, topological photonic routing is achieved in a bearded-stack VPC interface, originating from broadband unidirectional excitation of the valley-chirality-locked edge state by using a microdisk as a phase vortex generator. Control of valley in SOI platform not only**

**shows a prototype of integrated photonic devices with promising applications for delay line, routing, optical isolation and dense wavelength division multiplexing for information processing based on topological nanophotonics, but also opens a new door towards the observation of non-trivial states even in non-Hermitain systems.**

Silicon on insulator (SOI) provides a CMOS-compatible platform to fasten and enlarge data transfer both between and within silicon chips, by using optical interconnects to replace their electronic components[1]. Miniaturization of SOI devices can achieve highly-integrated photonic structures comprised of numerous optical components in a single chip, but increase inevitable backscattering that leads to energy loss and signal distortion. Consequently, backscattering suppression of light propagation is of fundamental interest and great importance for compact SOI integration. The discovery of topological photonics offers an intriguing way for robust information transport with pseudospin degree of freedom (DOF) of light[2, 3], particular for their capacities in backscattering-immune propagation and unidirectional transport. Such robustness is derived from the topological-protected edge states, enabling reflection-free transport between two topologically distinct domains[4-6], such as by using magneto-optical effect[2, 3, 7-10], 3D chiral structures[11-13], and bianisotropic metamaterials[14-16]. As a target to integrated topological nanophotonics, some all-dielectric strategies have been proposed recently. The first method to implement topological SOI structures was proposed by using array of coupling resonator optical waveguides at super-wavelength period[17], and have been exploited to topological-protected lasing effect[18, 19]. Applying all-dielectric photonic crystals (PCs) with crystalline symmetries to realize topological structures may reduce the size of devices[20]. Unidirectional coupling has been realized in a GaAs-based photonic crystal as a topological quantum optics interface[21]. These developments of topological nanophotonics open avenues to develop on-chip optical devices with built-in protection, such as robust delay lines, on-chip isolation, slow-light optical buffers, and topological lasers.

Valley pseudospin provides an additional DOF to encode and process binary information in two-dimensional transition metal dichalcogenide (TMDC) monolayers[22-24]. Analogous to valleytronics, exploration of valley in photonics renders powerful routes to address topological non-trivial phase by emerging an alternative valley DOF instead of photonic spin[25-30]. To retrieve topological valley phase, a general method is to break spatial-inversion (SI) symmetry for accessing opposite Berry curvature profiles near Brillouin zone corners, i.e. K and K' valley. Advanced in nanofabrication techniques, precise manufacture of the SI-symmetry-broken nanophotonic structures is easy to implement nowadays[31]. Consequently, valley photonic crystal (VPC) is a reliable candidate for SOI topological photonic structures, in particular for subwavelength strategy that still remains much challenge in topological nanophotonics. Furthermore, the topological valley phase below light cone ensures high-efficient light confinement in the plane of chip, such that photonic valley DOF naturally makes a balance between in-plane robustness and out-of-plane radiation. This is a crucial condition to design topological photonic structures for chips. Realization of topological valley transport in SOI platform is desirable for integrated topological nanophotonics.

In this work, we experimentally demonstrate a topological nanophotonic structure protected by valley phase at telecommunication wavelength. Our design is based on a standard SOI platform that allows integration with other optoelectronic devices on a single chip. Valley-dependent topological edge states can operate below light cone, benefiting to the balance between in-plane robust transport and out-of-plane radiation. Broadband robust transport is observed in sharp-turning interfaces constructed by two topologically-distinct valley photonic crystals, with footprint of 9× 9.2 μm$^2$. In addition, we have achieved topological photonic routing with high directionality, by exploiting unidirectional excitation of valley-chirality-locked edge state with a subwavelength microdisk.

In this work, our nanophotonic structures are prepared on SOI wafers with 220-nm-thickness silicon layers. As depicted in Fig. 1a, the valley photonic structure comprises two honeycomb photonic crystals (VPC1 and VPC2). The VPC layer is asymmetrically placed between SiO$_2$ substrate and top air region along z axis (see the inset of Fig. 1a). The upper panel of Fig. 1b gives the details of VPC1, where the unit cell (red lines) consists of two nonequivalent air holes, i.e. the smaller one $d_1$= 81 nm and the bigger one $d_2$ = 181 nm. On the other hand, the diameter of two air holes are altered to form another type of VPC (blue), i.e. VPC2 with $d_1$= 181 nm and $d_2$ = 81 nm. Here, VPC2 is the y-axis spatial-inversion-symmetry (ySI symmetry) partner of VPC1. Thus VPC1 and VPC2 have the same band structure, as shown in Fig. 1c. The details about VPC design can be seen in supplementary section A. Because the two air holes have different diameters that break the ySI symmetry, a bandgap (1360 nm ~ 1492 nm) emerges near the *K*(*K'*) valleys for TE-like polarizations.

Due to the bulk-edge correspondence, we will firstly study the bulk states of the first

TE-like band (labelled as 'TE1' in Fig. 1c), before discussing the topology of TE-like gap. The electromagnetic fields in the z-central plane (labeled as 'z = 0' in the inset of Fig. 1a) can mainly reflect the optical properties of VPC slab, so that we will focus on the field patterns at z = 0 plane in the following discussion. Take the eigenstates at K valley as examples. The simulated $H_z$ phase profile at z = 0 plane are plotted in Fig. 1d. We can see that the phase profile of VPC1/VPC2 increases anticlockwise/clockwise by $2\pi$ phase around the center of unit cell. Such optical vortex is related to valley pseudospin in electronic system, and thus can be termed as a photonic valley DOF. In TMDCs, the valley-polarized excitons can be selectively generated through control of the chirality of light[32, 33]. Similarly, the photonic valley is also locked to the chirality of excited light, i.e. left-circularly polarized (LCP) light couples to K'-valley mode while right-circularly polarized (RCP) light couples to K-valley mode[27]. To demonstrate this valley-chirality locking property remaining in SOI slab, we give the distribution of polarization ellipse of the in-plane electric field in Fig. 1e. Here, the polarization ellipse is generally defined as ellipticity angles[34] $\chi = \arcsin\left[2|E_x||E_y|\sin\delta / \left(|E_x|^2 + |E_y|^2\right)\right]/2$, where $\delta = \delta_y - \delta_x$ is the phase difference between $E_y$ and $E_x$. For VPC1, RCP response ($\chi = \pi/4$) exists in the singularity point of phase vortex at K valley (red center in the upper panel of Fig. 1e), and vice versa for VPC2 (blue center in the lower panel of Fig. 1e). Figure 1f shows the temporal evolution of the RCP and LCP responses, respectively. Such valley-chirality locking gives the possibility to manipulate photonic valley modes. For example, when we place a circular-polarized dipole source in the singular point of phase vortex, the photonic valley Hall effect (PVHE) can be observed (see supplementary section B).

The above observation of optical vortex and PVHE can be related to topologically non-trivial phase. Next, we numerically calculate the Berry curvature distribution and the corresponding topological invariant to insightfully confirm the topological valley phase in our proposed VPC slab[35]. See Methods for more details of numerical simulation. Figure 2b shows the Berry curvature of TE1 band for VPC1, calculated with the parallel gauge transformation[36]. The Berry curvature of VPC is mainly distributed near two valleys, i.e. singular sink at K valley while peak at K' (red line in the inset of Fig. 2b). On the contrary, VPC2 reverses the Berry curvature distribution of the two valleys (blue line in the inset of Fig. 2b). In general, the global integration of Berry curvature over the whole Brillouin zone, the so-called Chern number, is zero under the protection of time-reversal symmetry. Instead, the valley-dependent integration of Berry curvature gives rise to nonzero value, i.e. the valley-Chern index $C_K$ ($C_{K'}$). Thus we can use the valley Chern number $C_V = C_K - C_{K'}$ to characterize the topology of the whole VPC system, providing a new route to retrieve topologically nontrivial phase.

We should note that the VPCs open a large TE-like gap (~ 10%) to guarantee broad bandwidth operation. Thus the Berry curvature for both valleys will overlap with each other. As a consequence, the valley Chern number is not well-defined integer, i.e. $0 < |C_V| < 1$. And the edge dispersion will not gaplessly cross from the lower band to the upper band (see supplementary section C). Regardless of this side effect, the difference in the sign of valley Chern number will ensure the interface under the protection of valley-dependent topological phase, as long as the bulk state at K valley is orthogonal to K' valley. Therefore, such a novel design still enables broadband robust transport along ΓK/ΓK' direction against certain perturbations (such as sharp-bend corners or 10% random error of hole diameter), as the inter-valley scattering is suppressed due to the vanishing field overlapping between two valley states. As an intuitive example, we construct an interface by using two VPCs with opposite valley Chern number. As shown in Fig. 2a, the interface is bearded-shape stacked by the bigger holes. The upper domain (VPC1 in Fig. 1) has valley-Chern number $C_V < 0$, while the valley-Chern number of lower domain (VPC2 in Fig. 1) exhibits opposite sign ($C_V > 0$). The simulated patterns shown in Fig. 2d confirm that the propagating light at $\lambda = 1430$ nm will not suffer from backscattering even if it encounters sharp corners. Note that such backscattering-immune propagation is valid for other wavelengths inside the bandgap, maintaining optical broadband operation protected by topological non-trivial phase. In the next section, we will experimentally characterize this broadband robust transport phenomenon.

To experimentally demonstrate topological robust transport of the valley-dependent edge states, we employ advanced nanofabrication technique to manufacture the flat-, Z- and Ω-shape topological VPC interfaces. The SEM images of fabricated samples are shown in Fig. 2c. The devices were prepared on a SOI wafer, with a nominal 220 nm silicon layer and 2.0 μm buried oxide layer. After the definition of 370-nm-thickness positive resist through electron-beam lithography (EBL), inductively coupled plasma (ICP) etching step is applied to pattern the top silicon layer such that VPC structure and its coupling waveguide was formed. And then the resist was removed by using an ultrasonic treatment process. See Methods for more details of nanofabrication process. These processes are able to precisely achieve our designed structures even in close proximity (separation about 40 nm in the topological interface).

Next, we will characterize the broadband robust transport of the topological edge states at the Z/Ω-shape interface. The experimental setup is shown in Fig. S4. The TE-polarized continuous waves at telecommunication wavelength were coupled to the 1.7-µm-width input waveguide by using a polarization maintaining lensed fiber, and then launched into VPC sample from the left end of the topological interface. After passing through the VPC devices, the propagating wave was coupled to the output waveguide at the right end and then collected by another lensed fiber. The corresponding transmission spectra were were detected the signals by using an optical power meter, with tuning the operation wavelength of excited waves. Note that all the transmission spectra are normalized to the 1.7-µm-width silicon strip waveguide located in the same writing field near the VPC samples. See Methods for more details of optical characterizations. The left panel of Fig. 2e shows the measured transmission spectra in the wavelength range of 1320~1570nm, for flat-, Z- and Ω-shape topological interfaces. In the bandgap region (yellow), the spectra kept on the flat-top high-transmittance platform, even for sharp-bending geometry (green and red lines). This intriguing property indicates the broadband robust transport in the frequency interval from 1360 nm to 1492 nm, due to the topological protection in the absence of inter-valley scattering. Note that these experiment results, in good agreement with simulations (right panel of Fig. 2e), were measured in the compact structures with footprint $9 \times 9.2$ µm$^2$, verifying the SOI platform potentially enables to integrate many photonic components on a single chip, when the size of device is small enough. Benefit from subwavelength periodicity (about λ/4), the proposed SOI VPC can develop a high-performance topological photonic devices with a compact size (< 10 µm).

Unidirectional transport is another important property of topological photonic structures to manipulate the flow of light. We should emphasis that the realization of robust transport in the valley-dependent interfaces does not definitely correspond to unidirectional emission. For example, the zigzag-stack valley-dependent interfaces can not unidirectionally excite by using circularly-polarized source, due to protection of ySI symmetry around the center of the interface. The polarization ellipse of such odd/even-like edge state is predominantly linear at the positions where the light intensity is high. Therefore, breaking this symmetry is required to engineer the generation of vortex fields in the edge states. The above configuration (Fig. 2a) of bearded-stack interface has broken the ySI symmetry. As a consequence, the vortex fields around the bearded holes of the upper domain will interact with the lower domain, and thus generates chiral-flow edge states (see supplementary section C). As depicted in Fig. 3a, such chirality ensures the rightward (leftward) excitation by using RCP (LCP) source. Simulated results in Fig. 3b confirm that the proposed valley-dependent topological interface can realize unidirectionality through control of the source chirality. In fact, similar results have been studied in a glide-plane PC waveguide, by shifting one side of the waveguide by half a lattice constant[37, 38].

We should emphasize that the introduction of topological non-trivial phase can guarantee high directionality under chiral source excitation in relatively broadband operation, while the case of the topologically-trivial system always operates in a narrowband as it is sensitive to the source position with frequency variation. To quantitatively determine unidirectional coupling, we define the directionality for a given frequency as $\kappa_0 = (T_L - T_R)/(T_L + T_R)$, where $T_L$ and $T_R$ are the transmittances detected at the left and right end, respectively. Furthermore, we would like to analyze the global efficiency of unidirectional coupling inside the photonic bandgap, so that we define an averaged parameter related to the frequency-domain integration of $\kappa_0$ in the whole bandgap, i.e. $\langle \kappa_0 \rangle_{gap} = \left| \int_{gap} \kappa_0 d\omega \right| / \Delta\omega_{gap}$, where $\Delta\omega_{gap}$ is the bandwidth of photonic bandgap. $\langle \kappa_0 \rangle_{gap} = 1$ represent that the chiral source couples to pure left-/right-forward edge states for all frequency in the bandgap. Here we analyze the global directionality by focusing on two dominant factors, i.e., relative location ($D_x$ and $D_y$) and separation ($\delta_{Si}$) of silicon region between the two bearded holes (Fig. 3c). The phase map of Fig. 3d shows that the maximum $\langle \kappa_0 \rangle_{gap}$ emerges at the point of $D_x$ = 192.5 nm and $D_y$ = 111 nm, which is in correspondence with the valley-dependent topological interface. It is interesting that the proposed design based on valley topology can certainly find the point of high directionality, while the general method requires massive simulations, just like what we do in Fig. 3d.

On the other hand, considering a fixed relative position that $D_x$ = 192.5 nm and $D_y$ = 111 nm, the global directionality $\langle \kappa_0 \rangle_{gap}$ as a function of separation $\delta_{Si}$ is also retrieved in Fig. 3e, when tuning the diameters of bearded holes. We can see that $\langle \kappa_0 \rangle_{gap}$ will stand on a high-directionality platform (above 0.9), when the separation is less than 50 nm. Qualitatively, this is because such extreme separation will enhance the interaction of vortex fields between upper- and lower-domain bearded holes, and thus strengthen the valley-chirality coupling of the topological interfaces.

For unidirectional coupling of on-chip topological edge states, one may use chiral quantum dots under strong magnetic field at ultralow temperature[21, 39]. In this work, we aim to develop an all-optical strategy, for unidirectional excition of the valley-chirality locking edge states in the SOI platform. To do this, a subwavelength microdisk serving as a phase vortex generator[40], is introduced into the topological interface. Figure 4a shows the schematic of designed device, combining SOI VPC and microdisk. The fabricated sample around microdisk can be seen in Fig. 4b. There are two 373-nm-width strip silicon waveguides (label as 'WVG1' and 'WVG2') at the left of the sample. When incident light couples to the WVG1/WVG2 input waveguide, it will generate anti-clockwise/clockwise phase vortex at the designed-microdisk with a close-to-diffraction–limited scale (630 nm diameter). Due to valley-chirality locking, the edge state near K/K' valley can be selectively routed to the upper/lower topological interface, through control of the chirality of optical vortex inside microdisk. This shows a prototype of on-chip photonic routing device based on topological protection.

Far-field microscopy is used to verify the photonic valley-chirality locking property and topological routing effect. A 20X objective is used to predominantly collect the out-of-plane radiation from two non-uniform grating couplers (labelled as 'G1' and 'G2' in Fig. 4a) and then imaged by using an InGaAs CCD. For a given incidence waveguide, an asymmetric radiation is obvious between G1 and G2. For example, the microscope images are presented in Fig. 4d for the WVG1 incidence at $\lambda = 1400$ nm. In this case, the propagating light was routed to the upper interface and radiated from the G1 port. The asymmetry of photonic routing is reversed when the incidence port is flipped to the other interface (Fig. 4e). For comparison, we also fabricated a control sample that replaced the SOI VPC by two strip silicon waveguide (Fig. 4c). The near-equal routing profiles demonstrate low directionality in Figs. 4f and 4g. Valley-dependent undirectionality is already visible to be distinguished from the control experiment.

The intensity of each grating coupler was collected from CCD, and the intensities

$I_{G1}$ and $I_{G2}$ scattered from the upper (G1) and lower (G2) ports can be extracted with high signal-to-noise ratio, after subtracting the noise that mainly arises from background radiation of laboratory. The extracted intensities $I_{G1}$ and $I_{G2}$ reflect the amount of light emission that is coupled to the upper- and lower-propagating valley-dependent edge states, respectively. To experimentally qualify the directionality of routing devices, we define the directional coupling efficiency as $\kappa_{\exp} = (I_{G1} - I_{G2})/(I_{G1} + I_{G2})$. The full-band directional coupling efficiency can be measured by tuning the operation wavelength of the excited waves. Figure 4h shows the directionality spectra as a function of wavelength. In the bandgap, a strong and broadband directionality was observed. For anticlockwise-phase-vortex excitation, the incident light couples to valley-dependent topological edge states propagating along upper interface (red line in Fig. 4g). The directional coupling efficiency of the topological routing device is up to 0.5 within a broadband region due to topological protection. Note that the maximum $\kappa_{\exp}$ is up to ~0.895, implying 18:1 extinction ratio between G1 and G2. When the handedness of the excitation flips, so do the propagation directions of the valley-dependent topological edge states (blue line in Fig. 4h). For comparison, the low-directionality spectra for the control experiment were depicted in Fig. 4i. There are few discrete wavelength to reach $|\kappa_{\exp}| > 0.5$. A more experimental description is presented in the supplementary section E.

In summary, we have successfully applied the valley DOF to topologically manipulate the flow of light in silicon-on-insulator platform. Benefit from the below-light-cone operation, valley-dependent topological edge state is easy to balance in-plane robust transport and out-of-plane radiation. Topological robust transport and topological photonic routing are experimentally demonstrated and confirmed at telecommunication wavelength. Our study paves the way to explore photonic topology and valley in SOI platform, which is a promising system in taking advantage of topological properties into nanophotonic devices, particularly important for backscattering suppression and unidirectional transport. Furthermore, our subwavelength strategy enables to design compact-size topological SOI devices that allows integration with other optoelectronic devices on a single chip. It shows a prototype of on-chip photonic device, with promising applications for delay line, routing, optical isolation and dense wavelength division multiplexing for information processing based on topological nanophotonics. Finally, the platform of SOI topology opens a new door towards the observation of non-trivial states even in non-Hermitain photonic systems.

Note added. During preparation of this work, we are aware of a related work on experimental demonstration of valley-dependent edge states through air-bridge slab structures with sharp turning profiles[41].

**Numerical simulations**. In this work, all of the band structures and the corresponding eigenfield patterns were calculated by MIT Photonic Bands[42] (MPB) based on plane-wave expansion (PWE) method, while all of the optical transport calculations were implemented by MIT Electromagnetic Equation Propagation[43] (MEEP) based on finite-difference time-domain (FDTD) method.

**Figures and Figure Captions**

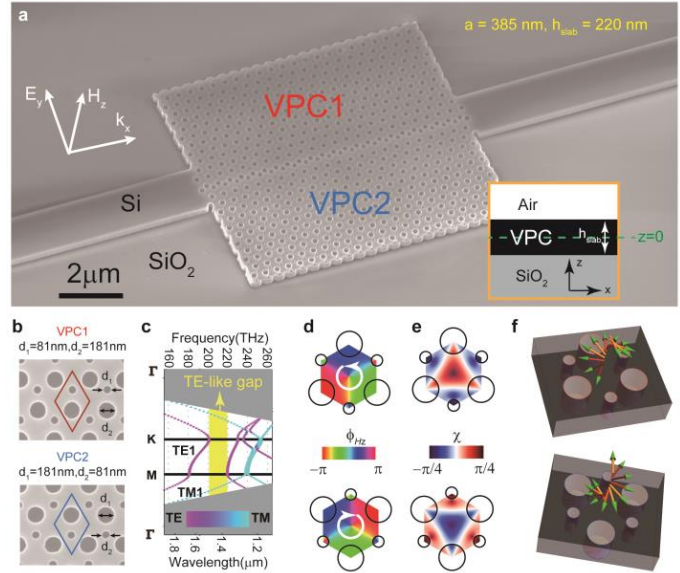

**Figure 1 | Band structures and nontrivial topology in silicon-on-insulator (SOI) valley photonic crystals (VPCs). a**, Oblique-view SEM of the fabricated VPC, which is patterned on standard 220-nm-thickness silicon wafer. Inset indicates the VPC slab is asymmetrically placed between SiO$_2$ substrate and top air region along z axis. **b**, Details of the unit cells consisting of two inequivalent air holes. **c**, Bulk band for SOI VPC. The colormap indicates the linear polarization of photonic band. Gray region: light cone of silica. **d**, Simulated phase vortex of H$_z$ field profile and **e**, ellipticity angle of (E$_x$, E$_y$) field at K valley of TE1 band for VPC1 (upper) and VPC2 (lower). **f**, Temporal evolution of RCP and LCP at the singularity point, respectively.

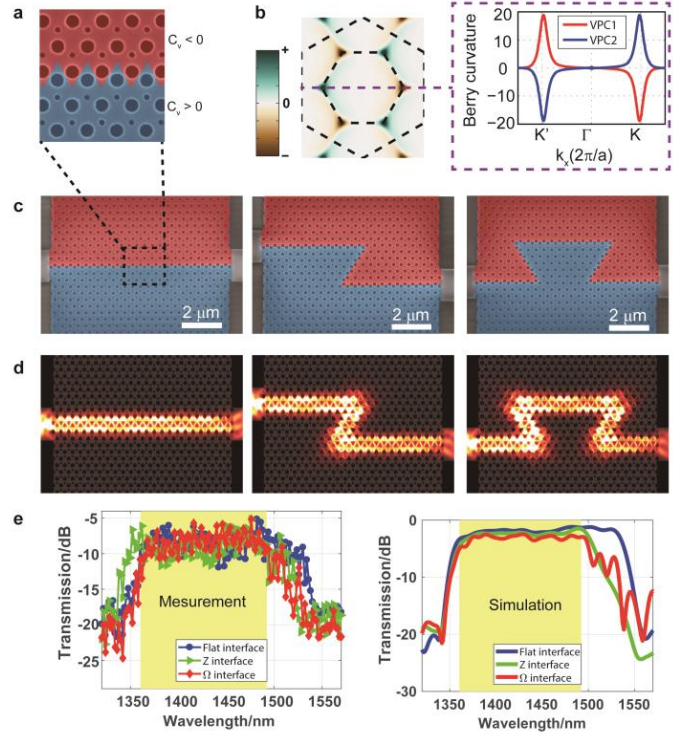

**Figure 2 | Valley-dependent topological edge states in SOI VPC. a**, SEM image of the valley-dependent topological interface constructed by two types of VPC slabs. **b**, Distribution of Berry curvature for TE1 band. Inset plots the Berry curvature along $k_y = 0$ direction, indicating that the Berry curvature of VPC2 (blue) has opposite distribution to that of VPC1 (red). **c**, Top-view SEM images of the fabricated samples, including flat, Z-shape and Ω-shape topological interfaces. **d**, Electromagnetic (EM) energy intensities for the three distinct shape topological interfaces at $z = 0$ plane ($\lambda = 1430$ nm). **e**, Measured and simulated transmission spectra for the topological flat (blue), Z-shape (green) and Ω-shape (red) interfaces, respectively. The experiment results show in good agreement with simulations.

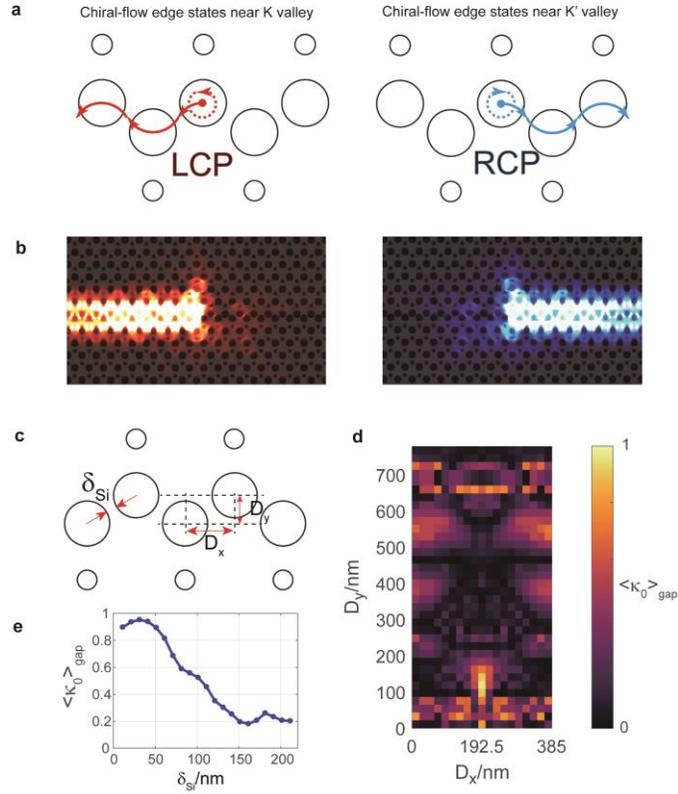

**Figure 3 | Valley-chirality-locked unidirectional excitation by using a circular-polarization chiral source. a**, Illustration of unidirectional coupling along the topological interface, depending on the selective excitation of phase vortex. **b,** Unidirectional propagation in bearded-stack interface by controlling the chirality of circular-polarized source at λ = 1430 nm. **c**, Key geometry parameters to affect directional coupling efficiency (DCE). Here, DCE is defined as $\kappa_0 = (T_L - T_R)/(T_L + T_R)$. **d**-**e**, Global analysis of DCE in the topological bandgap, under consideration of two dominant factors: **d**, relative location ($D_x$ and $D_y$) and **e**, separation ($\delta_{Si}$) of silicon region between the upper and lower bearded holes along the interface.

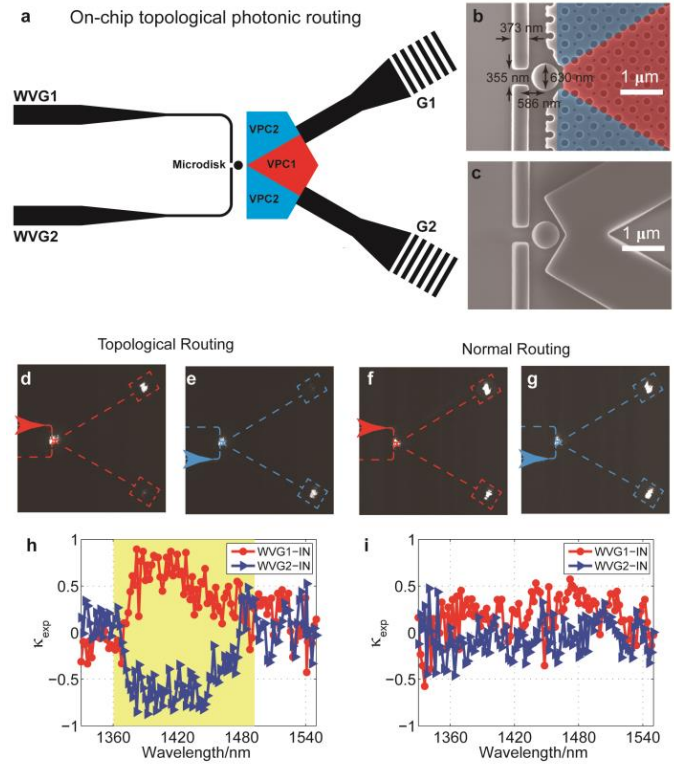

**Figure 4 | Experimental realization of topological photonic routing. a**, Schematic view of topological photonic routing sample, including the bearded-stack SOI-VPC interface and the microdisk. **b**, SEM of the sample near microdisk. **c**, Control sample with the same configuration to **b**, except for replacing the VPC by two strip silicon waveguides. **d-g**, Photonic routing profiles at λ = 1400 nm, imaged by an optical far-field microscopy (20X objective). **h-i**, Measured directionality spectra for topological (**h**) and normal (**i**) photonic routing devices. The spectra show the directional coupling efficiency, $\kappa_{exp} = (I_{G1} - I_{G2})/(I_{G1} + I_{G2})$, as a function of the operation wavelength.